\magnification 1200
\input amstex

\catcode`\X=11 \catcode`\@=\active
\documentstyle{amsppt}
\refstyle{A}

\NoRunningHeads

\catcode`\X=12\catcode`\@=11
\def\n@wcount{\alloc@0\count\countdef\insc@unt}
\def\n@wwrite{\alloc@7\write\chardef\sixt@@n}
\def\n@wread{\alloc@6\read\chardef\sixt@@n}
\def\crossrefs#1{\ifx\alltgs#1\let\tr@ce=\alltgs\else\def\tr@ce{#1,}\fi
   \n@wwrite\cit@tionsout\openout\cit@tionsout=\jobname.cit
   \write\cit@tionsout{\tr@ce}\expandafter\setfl@gs\tr@ce,}
\def\setfl@gs#1,{\def\@{#1}\ifx\@\empty\let\next=\relax
   \else\let\next=\setfl@gs\expandafter\xdef
   \csname#1tr@cetrue\endcsname{}\fi\next}
\newcount\sectno\sectno=0\newcount\subsectno\subsectno=0\def\r@s@t{\relax}
\def\resetall{\global\advance\sectno by 1\subsectno=0
   \gdef\firstpart{\number\sectno}\r@s@t}
\def\resetsub{\global\advance\subsectno by 1
   \gdef\firstpart{\number\sectno.\number\subsectno}\r@s@t}
\def\v@idline{\par}\def\firstpart{\number\sectno}
\def\l@c@l#1X{\firstpart.#1}\def\gl@b@l#1X{#1}\def\t@d@l#1X{{}}
\def\m@ketag#1#2{\expandafter\n@wcount\csname#2tagno\endcsname
     \csname#2tagno\endcsname=0\let\tail=\alltgs\xdef\alltgs{\tail#2,}%
   \ifx#1\l@c@l\let\tail=\r@s@t\xdef\r@s@t{\csname#2tagno\endcsname=0\tail}\fi
   \expandafter\gdef\csname#2cite\endcsname##1{\expandafter
     \ifx\csname#2tag##1\endcsname\relax?\else{\rm\csname#2tag##1\endcsname}\fi
     \expandafter\ifx\csname#2tr@cetrue\endcsname\relax\else
     \write\cit@tionsout{#2tag ##1 cited on page \folio.}\fi}%
   \expandafter\gdef\csname#2page\endcsname##1{\expandafter
     \ifx\csname#2page##1\endcsname\relax?\else\csname#2page##1\endcsname\fi
     \expandafter\ifx\csname#2tr@cetrue\endcsname\relax\else
     \write\cit@tionsout{#2tag ##1 cited on page \folio.}\fi}%
   \expandafter\gdef\csname#2tag\endcsname##1{\global\advance
     \csname#2tagno\endcsname by 1%
   \expandafter\ifx\csname#2check##1\endcsname\relax\else%
\fi
   \expandafter\xdef\csname#2check##1\endcsname{}%
   \expandafter\xdef\csname#2tag##1\endcsname
     {#1\number\csname#2tagno\endcsnameX}%
   \write\t@gsout{#2tag ##1 assigned number \csname#2tag##1\endcsname\space
      on page \number\count0.}%
   \csname#2tag##1\endcsname}}%
\def\m@kecs #1tag #2 assigned number #3 on page #4.%
    {\expandafter\gdef\csname#1tag#2\endcsname{#3}
    \expandafter\gdef\csname#1page#2\endcsname{#4}}
\def\re@der{\ifeof\t@gsin\let\next=\relax\else
    \read\t@gsin to\t@gline\ifx\t@gline\v@idline\else
\expandafter\m@kecs \t@gline\fi\let \next=\re@der\fi\next}
\def\t@gs#1{\def\alltgs{}\m@ketag#1e\m@ketag#1s\m@ketag\t@d@l p
    \m@ketag\gl@b@l r \n@wread\t@gsin\openin\t@gsin=\jobname.tgs \re@der
    \closein\t@gsin\n@wwrite\t@gsout\openout\t@gsout=\jobname.tgs }
\outer\def\localtags{\t@gs\l@c@l}
\outer\def\globaltags{\t@gs\gl@b@l}
\outer\def\newlocaltag#1{\m@ketag\l@c@l{#1}}
\outer\def\newglobaltag#1{\m@ketag\gl@b@l{#1}}

\def\t@gsoff#1,{\def\@{#1}\ifx\@\empty\let\next=\relax\else
\let\next=\t@gsoff
   \expandafter\gdef\csname#1cite\endcsname{\relax}
   \expandafter\gdef\csname#1page\endcsname##1{?}
   \expandafter\gdef\csname#1tag\endcsname{\relax}\fi\next}
\def\verbatimtags{\let\ift@gs=\iffalse\ifx\alltgs\relax\else
   \expandafter\t@gsoff\alltgs,\fi}
\def\(#1){\edef\dot@g{\ift@gs\if@lign(\noexpand\etag{#1})
    \else\ifmmode\noexpand\tag\noexpand\etag{#1}%
    \else{\rm(\noexpand\ecite{#1})}\fi\fi\else\if@lign{\rm(#1)}
    \else\ifmmode\noexpand\tag#1\else{\rm(#1)}\fi\fi\fi}\dot@g}
\let\ift@gs=\iftrue\let\if@lign=\iffalse
\let\n@weqalignno=\eqalignno
\def\eqalignno#1{\let\if@lign=\iftrue\n@weqalignno{#1}}

\catcode`\X=11 \catcode`\@=\active

\localtags
\TagsOnRight


\define\R{\Bbb R}
\define\real{\Bbb R}
\define\reald{\Bbb R^d}
\define\eps{\epsilon}
\define\sgm{\sigma}
\define\tsgm{{\tilde \sigma}}
\define\trho{{\tilde \rho}}
\define\tmu{{\tilde \mu}}
\define\tJ{{\tilde J}}

\define\rhe{\rho ^\epsilon}
\define\goes{\rightarrow}
\define\Z{\Bbb Z}

\vskip.5truein
\topmatter
\title
Phase Segregation Dynamics\\
In Particle Systems\\
with Long Range Interactions II:\\
Interface motion.
\endtitle

\leftheadtext\nofrills
{G.Giacomin\\
Institute of Applied Mathematics\\
University of Zurich\\
Winterthurer Str. 190\\
CH--8057 Zurich, Switzerland\\
gbg@amath.unizh.ch\\
 and\\
 J.L.Lebowitz\\
Departments of Mathematics and Physics\\
Rutgers University\\
New Brunswick, NJ 08903\\
lebowitz@math.rutgers.edu}
\rightheadtext\nofrills
{Interface motion}

\author
Giambattista Giacomin and Joel L. Lebowitz
\endauthor
\affil
Universit\"at Z\"urich and Rutgers University
\endaffil
\address
Giambattista Giacomin
\hfill\newline
Institut f\"ur  Angewandte Mathematik
\hfill\newline
Universit\"at Z\"urich--Irchel
\hfill\newline
Winterthurerst. 190, CH--8057 Z\"urich Switzerland
\endaddress
\email
gbg\@amath.unizh.ch
\endemail
\address
Joel L. Lebowitz
\hfill\newline
Department of Mathematics and Physics
\hfill\newline
Hill Center, Rutgers University
\hfill\newline
New Brunswick, N.J. 08903, U.S.A.
\endaddress
\email
Lebowitz\@math.rutgers.edu
\endemail
\thanks
{\it Short title:} Phase segregation with Kac potentials
\endthanks
\abstract
We study  properties of the solutions of a
 family of second order integro-differential equations, which
 describe the large scale
dynamics of  a  class of microscopic
phase segregation models
with
particle conserving  dynamics.
We first establish existence and uniqueness as well as some
properties of the instantonic solutions.
Then we
concentrate on  formal asymptotic (sharp interface) limits.
We argue
that the obtained interface evolution laws
(a Stefan--like problem and  the Mullins--Sekerka solidification
model)
coincide with the ones which can
be obtained in the analogous limits from the
Cahn--Hilliard equation, the fourth order PDE
which is the standard macroscopic model for
phase segregation with one conservation law.
\endabstract
\endtopmatter

\document
\resetall

\head
1. Introduction
\endhead
\sectno=1

In part I [\rcite{GL1}]
 we rigorously
derived  a  macroscopic equation describing the
time evolution of the empirical average process, i.e.
the local density, for a model of interacting particles
with one conservation law undergoing phase segregation.
Here
we establish  several properties of the solutions of a class
of such equations, which
are expected to describe more general microscopic models
(see Section 3 of [\rcite{GL1}]).  Our emphasis is
 on the late stages of phase segregation,
when the phenomena are dominated by the motion of sharp interfaces
separating
domains of different densities.

Let us briefly recall in an informal way
 the results in part I.
The particle models  are dynamic versions
of lattice gases with long range Kac potentials.
By a long range Kac potential, we mean
that the interaction energy between two particles, say between a particle at
$x$ and one at $y$ ($x$ and $y$ are both in
${\Bbb Z}^d$), is given by $\gamma ^d J(\gamma (x-y))$,
where $J$ is a smooth compactly supported function  ($J(r)=
J(-r)$) and $\gamma$ is a positive
parameter which is sent to zero.
The equilibrium states for these models have been
investigated thoroughly ([\rcite{Kac}],[\rcite{LP}], [\rcite{PL}])
and have provided great mathematical insight
into the static aspects of  phase transition phenomena.
The  dynamical version  of these
 models, in which each particle  jumps at random
times from a site of the lattice $\Z ^d$
to one of its unoccupied nearest neighbor sites is  sometimes called {\sl
local mean field Kawasaki dynamics}.  
The jump times are  chosen according
to a probability distribution which depends on the
particle configuration and is reversible with respect to
the equilibrium Gibbs measure.
To find a hydrodynamic scaling limit, we scale also the lattice spacing
with $\gamma $ and the time with $\gamma ^{-2}$ (this is the so
called {\sl diffusive limit}).
We then derive a (deterministic) evolution law for the
macroscopic density $\rho$.

In [\rcite{GL1}] we argue, but do not
prove, that our results extend to the case in which
there is a local (i.e. short range) interaction between the particles
in addition  to the Kac potential: the local interaction
should however be sufficiently weak, more precisely the system
with the local interaction {\sl alone} should be in
a high temperature regime. In the present paper we shall
consider the equations expected in this more general case.
They are given in terms of
 a second order integrodifferential equation:
$$
\partial _t\rho(r,t)=
\nabla\cdot \left[
\sigma(\rho) \nabla\left(
{\delta {\Cal F} \over \delta \rho}\right)
\right]
\(maineqpre)
$$
 $\rho$ is the density,
$\sigma$ (a function of $\rho$) is the mobility and
$$
{\Cal F}(\rho)=
\int_{T^d} f_c(\rho(r)) \text{d}r +{1\over 4}
\int \int _{T^d\times T^d}
J( r- r^\prime  ) \left[\rho (r) -\rho (r^\prime)\right]^2
\text{d} r \text{d} r^\prime
\(newway)
$$
in which $f_c$ is either convex  or it has  a symmetric double well
structure,
with minima at $\rho^-$ and $\rho^+$.  The latter will be the
relevant case for us: we will call
the densities $\rho^-$ and $\rho^+$ the {\sl phases}
of the system. The dependence on the temparature (of $\sigma$ and $f_c$)
is suppressed.
In probabilistic terms, \(maineqpre) is a law of large numbers for the
empirical averages over the particle system.
Equation
\(maineqpre) is in a particularly enlightening form:
it is the gradient flux
associated with the classical local mean field free energy functional
\(newway)
([\rcite{LP},\rcite{PL}]), with a density dependent mobility
$\sigma$.
The form of equation
\(maineqpre)  allows us to  connect the concepts of
  critical temperature,  phases,
 stable, unstable and metastable states of the model,
with the properties of the solutions of the
evolution equation. The next step is to understand
the evolution of the boundaries (interfaces) between regions
in which the density is close to $\rho^\pm$.

Formal matched asymptotic
expansions, in the so called
{\sl sharp interface limit}, of the solution of  macroscopic evolution
equations
(see e.g. [\rcite{Langer},\rcite{Pego},\rcite{Caginalp},\rcite{CENK}]) have
been successfully
employed to understand the interface motion in
several models.
By {\sl sharp interface} we mean the limit in which the
phase
domains are very large with respect to the size of the interfacial region:
if we denote by $L$ the {\sl typical }
size of the domains, we will look for results in the limit $L \rightarrow
\infty$. The time will have to be properly scaled as well, typically
as some integer power of $L$, according to the type
of initial condition
(see Section 3).

The general picture that we obtain for the interface motion
is the following:  
choose an initial condition
which takes only metastable or stable values over large
 domains(of typical diameter
$O(L)$), while  the interfacial regions are layers of thickness
$O(1)$ and let it evolve according to \(maineqpre).
There is, first, the equilibration of the interface which happens
on a fast time scale ($t \ll L^2$). Then, on the time scale
$t \propto L^2$ the evolution of the density in the bulks
(that is the interior of the domains) is given by a nondegenerate
nonlinear diffusion equation with Dirichlet boundary conditions
on a free boundary, the interface ({\sl Stefan problem}). Once the density
in the bulks is
 relaxed to the density of the phases, the
motion of the interface on this time scale stops. A slower
evolution can then  be seen on the time scale $t \propto L^3$ and
the motion of the interfaces is given by the Mullins--Sekerka
model; a quasistatic free boundary problem in which the mean curvature
of the interface plays a fundamental role.

Throughout the paper we will repeatedly
stress the fact that the evolution in the sharp interface
limits we have just outlined are the same as those
obtained from the Cahn--Hilliard equation in the corresponding limits.
The Cahn--Hilliard equation
[\rcite{CH}] is an equation of the form \(maineqpre)
but the free energy functional is chosen of
the Ginzburg--Landau type
$$
{\Cal F}_{GL} (\rho)
=
\int \left[
V(\rho) +\vert \nabla \rho \vert^2
\right] \text{d}r
\(GLfree)
$$
in which $V$ is a smooth function
(say a polynomial) with a double
minimum
since the aim is to study phase segregation phenomena.
The results on sharp interface motion for the Cahn--Hilliard equation have
been obtained
formally by R. L. Pego [\rcite{Pego}] and part
of his program has been 
made rigorous recently [\rcite{ABC},\rcite{BK}]. In  [\rcite{Pego}]
only the case $\sigma=1$ is considered, but the
general case is analogous as long as the mobility does not
vanish in  the phases (the minima of $V$).
The differences between the two
models in this case are mostly  in their having
different surface tension, which appears as a factor in the formulas:
the parameters of the models can therefore be  tuned to give
exactly the same sharp interface motion.
A possibly  deeper difference arises in the case of zero temperature
(in our case the temperature is built in the model while in the
Cahn--Hilliard case one has to choose an appropriate $V$ which
is temperature dependent, see e.g. [\rcite{CENK}]):
the profile of the stationary front solutions ({\sl instantonic solutions})
in the two cases has  a substantially different limit.
 The instantonic solution  in our case
approaches a step function, while in the C--H case it
approaches a differentiable function. The zero temperature case
has been treated  (formally) in [\rcite{CENK}],
but here we will limit ourselves to
a few observations in this case.


Finally we remark that in the case of  particle dynamics without
conservation law (Glauber or Kawasaki+Glauber dynamics
[\rcite{Spohn}]) there are by now several results proving that
 the interface  motion is by mean curvature
([\rcite{Bonaventura},\rcite{DOPT},\rcite{KS},\rcite{Spohninterface}]).

The paper is organized as follows:
in Section 2 we introduce in an informal way the equations and all the relevant
thermodynamical concepts. In Section 3 we introduce the
sharp interface limit and state the results
of the formal asymptotics: the computations are carried out in Section 5.
In
 Section 4 we state and prove
existence and uniqueness for the equations introduced in Section 2
as well as a result of existence and uniqueness of
{\sl istantonic} solutions.

\head
2.The nonlocal equation
\endhead
\resetall
\sectno=2

In this Section we introduce the evolution equations in an informal
way, assuming that all the functions which appear in the formulas are suitably
smooth.
Moreover, in order to understand better the role of the quantities appearing
in the definitions, we will restrict ourselves for the moment to the cases
in which we will study phase segregation phenomena.
 Many of the assumptions made here
are not needed to establish properties like existence
and uniqueness: these will be stated and proven in a more general
context in
Section 4.

Let $T^d= \real^d$~mod~$a {\Bbb Z}^d$ be the torus of diameter
$a\in \real^+$ (seen as a metric space, equipped with the
periodic Euclidean distance).
Occasionally we will consider the case $a=+\infty$, that is $T^d=
\real^d$, but, unless esplicitly stated, $a$ has to be considered fixed
(and finite).

\definition{The free energy and the mobility}
We start by introducing the free energy
functional
$$
{\Cal F}(\rho)=
\int_{T^d} f(\rho(r)) \text{d}r -{1\over 2}
\int \int _{T^d\times T^d}
J( r- r^\prime  ) \rho (r) \rho (r^\prime)
\text{d} r \text{d} r^\prime
\(freeE)
$$
in which  $\rho$ is a function from $ T^d$ to $[0,1]$,
$f:[0,1]\rightarrow \R$
is convex and we will refer to  $f(\rho)$ as the
{\sl free energy of the reference system} corresponding to the 
{\sl interaction potential} $J$ being zero.  $J$  is chosen {\sl attractive},
$$
J\ge 0, \(extra1)
$$
{\sl isotropic},
$$
J(r) \text{ depends
only on } \vert r \vert
\(extra2)
$$
and  is compactly supported in a
 ball
of diameter smaller than the diameter of $T^d$.

Define moreover the {\sl mobility} $\sigma$, a function
from $[0,1]$ taking nonnegative values and such that
$\sigma(0)=\sigma(1)=0$.
\enddefinition

\definition{Further assumptions on $f$ and $\sigma$}
We will make  three further assumptions on $f$ and
$\sigma$:

\noindent
1. {\sl Symmetry.} We assume
that both $f$ and $\sigma$ are symmetric with respect to $1/2$,
i.e.
$$
f((1+m)/2)=f((1-m)/2), \ \ \  \ \ \
\sigma((1+m)/2)=\sigma((1-m)/2)
\(assume4)
$$
for all $m \in [-1,1]$. This is a consequence
of the particle--hole symmetry in the microscopic dynamics.

\noindent 2. {\sl The reference system is a nondegenerate diffusion.}
We will require that there exists $c\in [1, \infty)$ such that
$$
{1\over c }\le D(\rho)\equiv\sigma(\rho) f^{\prime \prime}(\rho)\le c
\(nondegen)
$$
for all $\rho\in (0,1)$. We will assume that the limits
of $D(\rho)$ when $\rho$   approaches $0$ and $1$ exist
and these will extend the definition
of $D(\cdot)$ to $[0,1]$.

\noindent
3. {\sl There are at most two phases.}
If we set
$$
f_c(\rho)=- {\hat{J}(0)\over 2}\left(
\rho -{1\over 2} \right)^2
+
f(\rho)
\(fbeta)
$$
where $\hat{J}(0)=\int J(r) \text{d}r$,
 the free energy
\(freeE) becomes,  up to a irrelevant additive constant, the
functional
defined in \(newway).
If $\rho$ is constant the second term in
\(newway) vanishes, so that $f_c$
is the  free energy density of a homogeneous profile
(or {\sl constrained equilibrium} free energy, see
[\rcite{LP}, \rcite{PL}]).
For (small enough) values  of $\hat{J}(0)$, $f_c$ can
be convex: in this case there is no phase segregation.
We are then focusing on the nonconvex case and
we assume that $f_c$  has a double
well structure:
precisely,
there exists $\rho_m^+,\, \rho_m^-\in (0,1)$, $2\rho_m^+-1=
1-2 \rho_m^-$, such that
$$
\cases
f^{\prime\prime}_c(\rho)<0, &\text{if }\rho \in (\rho_m^-,\rho_m^+)
\\
f^{\prime\prime}_c(\rho)>0, &\text{if } \rho \in (0,\rho_m^-) \cup
(\rho_m^+,1).
\endcases
\(extra3)
$$
Once $f_c$ is non convex, assumption 
\(extra3) is, for example,  consequence of the stronger
assumption
$$
f^{\prime\prime} (\rho)
\text{ is strictly increasing for }
\rho >1/2
\(GHS)
$$
which holds if the reference system is ferromagnetic
(see Section 1 of [\rcite{Joel}]).

\enddefinition

\definition{The evolution problem} We consider the equation
$$
\partial_t \rho=
\nabla \left[
\sigma \nabla
\left(
{\delta {\Cal F} \over \delta \rho }
\right)
\right]
\(maineq)
$$
with initial condition  $\rho(\cdot, 0)=\overline{\rho}(\cdot)$.
An alternative way
to write \(maineq) is
$$
\partial_t \rho (r,t)=
\nabla \left[ D(\rho(r,t)) \nabla \rho(r,t) -\sigma (\rho(r,t))
(\nabla (J(\cdot)*\rho(\cdot,t))(r)
\right]
\(spelledout)
$$
where $*$ denotes the convolution in $T^d$. Global existence and
uniqueness  for a weak version of
\(spelledout) is given in  Theorem 4.1; see  Remark 4.1
for the regularity properties.

\enddefinition

\definition{Stable, unstable and metastable states}
 By assumption \(extra3)
the interval $(0,1)$ can be partitioned into
three sets:

\noindent
(1) the {\bf unstable region}
$$
[\rho_m^-, \rho_m^+]
\(unstable)
$$
in which $f_c^{\prime\prime}\le 0$ and $\rho^\pm_m$ are
defined by $f_c^{\prime\prime}
(\rho_m^\pm)=0$;

\noindent
(2) the {\bf metastable region}
$$
[\rho^-,\rho_m^-)\cup (\rho_m^+, \rho^+]
\(metastable)
$$
where $\rho^\pm$ are defined by
$f_c^\prime (\rho^\pm)=0$;

\noindent
(3) the {\bf stable region}
$$
[0,\rho^-)\cup (\rho^+, 1]
\(stable)
$$
in which $f_c^{\prime\prime}$ is bounded away from zero (and it is
$+\infty$ at $0$ and $1$).
The two minima $\rho^\pm$ are called {\sl phases}.
\enddefinition

All the definitions and assuptions  we have made are consequences of the
statistical mechanical origin of the model and they are explained
at length in [\rcite{GL2}] and [\rcite{GL1}]. In particular the fact that
the density $\rho$  is between 0 and 1 follows from the
 original particle model  which cannot have more
than one particle at  a lattice site.
Another  consequence of this is  that
 $\sigma(0)=\sigma(1)=0$. The free energy $f$
comes from a  particle model which is above its critical temperature and
it is then a smooth function in $(0,1)$ [\rcite{Simon}]. The
control on the regularity of the mobility $\sigma$
is  harder, but Lipschitz continuity (see  \(assume3))  can be shown
to hold in a rather general context [\rcite{Spohn}].
The assumption \(nondegen) is natural, since
$D$ is the diffusion coefficient of a high temperature system
(see [\rcite{GL1}] and [\rcite{Spohn}]).
We stress however that there is no  fundamental reason to
assume attractivity and isotropy of  $J(r)$ to
derive \(maineq).
Even in a non attractive
case interesting phase segregation phenomena might occur:
this is not considered in this paper. Moreover, in the anisotropic case one
can do a formal asymptotic
analysis as in Section 5 and get more general results
than the ones we present;
 see \(assume2) for a condition which replaces the isotropy.

\definition{A particular case}
For the  particle system treated in [\rcite{GL1}] with complete
proofs
we have free energy density
$$
f(\rho)={1\over \beta}(\rho \log \rho +(1-\rho) \log(1-\rho))
\(entropy)
$$
and  mobility
$$
\sigma ^0(\rho)=\beta\rho (1-\rho)
\(sigma)
$$
where $\beta$ is a positive parameter, corresponding to the inverse
of the  temperature.
In this case, for $\beta \hat{J}(0) > 1/4$, $f^{\prime\prime}_c(1/2)<0$,
while for $\beta \hat{J}(0)<1/4$ we have that
$f_c^{\prime \prime} (\rho)>0$ for
all $\rho \in (0,1)$.
Notice that in this example $D=1$.
\enddefinition

\head
3. The sharp interface limit
\endhead
\resetall
\sectno=3

We want to study the motion of boundaries between regions 
in which the density  is
either stable or metastable. Our analysis is based on formal matched
asymptotic expansions.

\definition{Space--time rescaling}
in what follows we shall consider the evolution
\(maineq) with $r $ in $ \epsilon ^{-1}T^d$ with
$\epsilon >0$  and we will scale time as $\epsilon^{-q}$, with $q=0, 2,3$
(only a short remark will be made
about $q=4$), according to the case.
We will make the substitution $t=\epsilon ^{-q}\tau_q$.
At this stage, it is
more convenient to reformulate
the problem in terms of the rescaled densities
$$
\rhe (r, \tau _q )=
\rho  (\eps ^{-1} r, \eps ^{-q} \tau _q)
\(rhoeps)
$$
in which the spatial variable $r$ ranges in $T^d$ and $\tau_q$ is a positive
number.
The evolution equation \(maineq) yields
$$
\eps ^{q-2} \partial _{\tau _q}\rhe (r, \tau _q)
=\nabla \left[
\sigma (\rhe (r ,\tau _q)) \nabla
 \mu ^{\eps}(r, \tau _q)
\right]
\(mainresc)
$$
in which
$$
\mu^\eps (\rhe)(r, \tau) =
f^\prime(\rhe (r, \tau ))
- \int _{T^d }
J_\eps ( r- r^\prime  ) \rhe (r^\prime, \tau )
\text{d} r^\prime
\(mueps)
$$
and $J_\eps (r)= \eps ^{-d} J (\eps ^{-1} r)$. Recall that
$J$ is taken to be supported in a fixed ball,
independent of $\eps$.
\enddefinition

Before starting the analysis of the time evolution of the system, we need
to introduce the concept of instantonic solution.

\definition{The one dimensional problem and its instantonic solution }
We replace  $T^d$ with
$\real ^d$ in \(maineq) and let
the initial condition $\rho ^0(r)=u_0(r_1)$ depend  only on
the $1-$coordinate. Then \(maineq) reduces to the
one dimensional problem $(r_1=z$)
$$
\partial _t u (z,t)=
\partial_z \left[\sigma (u(z,t))\left(
\partial _z f^\prime (u(z,t)) - \partial_z ((\tilde J*u) (z,t)\right)
\right]
\(main1d)
$$
in which
$$
\tilde {J} (z)=
\int_{\real ^{d-1}}
J(z, r_2,\ldots ,r_{d}) \text{d}r_2 \ldots
\text{d}r_{d}
\(Jtilda)
$$
The initial condition is $u(\cdot,0)=u_0(\cdot)$.
Clearly
$u(z,t)\equiv C\in[0,1]$ is a stationary solution.
It is easy to see that,  if $C$ is in the unstable region, this
solution is unstable. In Section 4, Proposition 4.2, we show
that \(main1d) admits a
 nonconstant stationary solution which is unique up
to translation and reflections in a proper class (see Proposition 4.2
for the precise statement).
  $U$ solves the functional equation (see the proof of Proposition 4.2)
$$
U (z)
= (f^\prime)^{-1}
\left( \left(\tilde J  *\left(U -{1\over 2}\right)(z)\right)
\right)
\(functional)
$$
in which $(f^\prime)^{-1}$ is the inverse function of
$f^\prime$, which is invertible since we
assumed $f^{\prime \prime}>0$.
  From \(functional) it follows directly that
$$
\lim_{z\rightarrow \pm \infty}
U(z)= \rho^\pm
\(asymptph)
$$
hence the asymptotic values for $U$,  which we will
 call the instantonic solution,
are exactly the phases.
We observe that
$\rho (r,t) =U (z_0 +\nu \cdot r)$ is a stationary
solution of \(maineq) in the case $T^d =\reald$
for any choice of $z_0 \in \real$ and $\nu \in \reald$,
$\vert \nu \vert =1$. This solution represents a plane stationary
front.
\enddefinition

\definition{The sharp interface initial condition}
The type of initial density profiles $\rhe_0$ that we wish to
consider take values which are either metastable or stable
on the whole space  $T^d$, with the exception of the points which are
at distance $O(\epsilon )$ from a smooth (hyper)surface
$\Gamma_0 \subset T^d$, which will be called the   interface.
We will denote by $\nu(r)$ one of the two unit vectors
normal to the interface at the point $r$: the (conventional)
choice of the orientation will be explained below.
We suppose that
there exists a function $\rho_0$, defined in $T^d$ and smooth outside
 $\Gamma _0$, such that if $r$ is not on the interface, then
$$
\lim_{\epsilon \rightarrow 0} \rhe_0(r)= \rho_0(r)
\(shrpinit)
$$
and the corresponding limit holds also for the derivatives of  $\rhe_0$.
Moreover if $\phi (r, \Gamma _0)=O(\epsilon)$
($\phi$ is the signed distance), then
$$
\rho_0^\epsilon (r)=
U (\epsilon ^{-1} \phi (r, \Gamma _0))
+O(\epsilon)
\(extra)
$$
In particular the limit of
$\rho_0(r)$ as $r$ approaches $\Gamma _0$ from the interior
of a  cluster is
either $\rho^+$ or $\rho ^-$ and $\rho_0$ is discontinuous
at $\Gamma _0$. The sign of $\phi(r, \Gamma_0)$ is chosen to be
positive (respectively negative) if $\rho_0(r) >1/2$
(respectively $\rho_0(r) <1/2$) and $\nu $ is chosen to point
in the direction in which $\phi$ increases, i.e.
$\nu =\nabla \phi$ on the interface.

It turns out that the
 relevant time scale for the evolution of $\Gamma _0$
is $t \propto \epsilon ^{-2}$,
that is $q=2$ in \(mainresc).

\enddefinition

{\bf A remark about the equilibration of the interface.}
The sharp interface initial condition that we introduced
is such that the interface is locally stationary, since it
is locally very close to the  plane stationary front.
It is however very natural to consider initial condition
which satisfy \(shrpinit), but not necessarily
\(extra). In particular, the limiting
valuess of $\rho_0$ on the interface may not
coincide with the phases.
In this case, the
 first observation is that $\partial _t \rho ^\epsilon (r,t)=O(\epsilon ^2)$
for $t=0$ and $r$ away from the interface.
However if  $r$
is close to the interface then
$\partial _t \rho ^\epsilon (r,t)=O(1)$ at $t=0$.
We then expect the transition layer around the
interface to evolve on the time scale $O(1)$
and eventually (asymptotically on the time scale $O(1)$) to
 approach a profile which, to leading order, is given by $U
(\epsilon ^{-1} \phi (r, \Gamma_0))$.
This is a strong {\sl ansatz} on the behavior of the solutions
of \(maineq).
Due to the conservation of mass, an immediate problem arises:
the two asymptotic values of $U$, i.e. the two phases, will
not  in general match with the solution away from the interface.
This happens whenever the values of $\rho_0$ near $\Gamma_0$
do not coincide with the phases. In this case we have to deal
with a boundary layer problem.
 This early stage of the evolution can be studied,
at least  on a formal level,  and it gives rise
to a self similar Stefan problem: the formal expansion is
absolutely identical to that in Section 3 of [\rcite{Pego}]).
 We will not focus on this  stage
of the evolution and we will only remark that the boundary layer
should be of thickness $O(\epsilon \sqrt{t})$, so that it disappears
if $t=O(\epsilon ^{-2})$.
 We will focus on the later stage
by assuming that this stabilization
has taken place.

{\bf Sharp interface limit I: Stefan problem ($q=2$).}
If we assume that the interface is stable (see [\rcite{ABC}] for
some rigorous statements on stability of interfaces for the Cahn--Hilliard
model)
the structure of the initial condition is essentially preserved
 and hence $\rho^\epsilon(r, \tau_2)$
(together with its derivatives)
converges  to $\rho(r, \tau_2)$ away from a hypersurface $\Gamma _{\tau_2}$
 while the boundary values of $\rho(r, \tau_2)$ at $\Gamma$
coincide with the phases. Moreover we assume that $\rho$ is smooth
outside $\Gamma_{\tau_2}$ and that $\Gamma_{\tau_2}$ is itself smooth and
$(d-1)$--dimensional. In general this will be true only for $\tau_2$
in a finite interval and when a singularity appears one should introduce
a weak version of the evolution law.
We have the natural domain decomposition
$$
T^d=\Gamma_{\tau_2} \cup \Omega _{\tau_2}^+ \cup \Omega _{\tau_2}^-
\(natdecomp)
$$
and the sets appearing on the right--hand side of
\(natdecomp) are disjoint.

\proclaim{Statement 1} In the setting specified above
and at the level of formal asymptotics,
$\rho^\epsilon (r ,\tau_2)$ converges to $\rho (r, \tau_2)$,
which solves the following nonlinear Stefan problem
$$
\cases
\partial _{\tau_2} \rho =
\nabla \cdot\left(\sigma (\rho)\nabla \mu _0(\rho )\right)
& \text{ for }  r\in \Omega ^{\pm}_{\tau_2}\\
\lim _{r ^\prime \rightarrow
 (r)^\pm}
\rho (r^\prime , \tau _2)=\rho ^\pm
& \text{ for } r \in \Gamma _{\tau_2}\\
\rho(r,0)=\rho_0(r) &\text{ for all }r \in T^d\\
\endcases
\(stefan)
$$
in which $\mu_0 (\rho)= f^\prime _c (\rho ) $
and by limit as $r^\prime$ goes to $(r)^\pm$ we mean
the limit $\lambda \rightarrow 0^\pm$ with 
$r^\prime=r+\lambda \nu$.
The motion of the interface is generated by the velocity field
$V_2(r) \nu(r)$, where $r\in \Gamma _{\tau_2}$ and
$$
V_2(r)= { \sigma^\pm \over (2\rho ^+ -1)}
\left[ \nu \cdot \nabla  \mu _0 \right]^+_- (r)
\(vel2)
$$
in which $[ \nu \cdot \nabla  \mu _0 ]^+_- (r)=
\lim _{r^\prime \rightarrow
(r)^+} \nu (r) \cdot\nabla \mu_0(r^\prime)-
\lim _{r^\prime \rightarrow
(r)^-} \nu (r) \cdot \nabla \mu_0(r^\prime)$,
i.e. the jump in the normal gradient of $\mu_0$ across
the interface and
$\sigma^\pm=\sigma (\rho ^+)=\sigma (\rho ^-)$.
\endproclaim

{\it Remark 1:} in the case $\Gamma _0=\emptyset$, that is
the density takes only stable or metastable values
and there is no interface, this result has been  made rigorous,
even starting directly from the particle system
(see references in [\rcite{GL1}]).

{\it Remark 2:}
The Stefan problem we just introduced is expected to stabilize
as $\tau_2 \rightarrow \infty$ and it should lead to
homogeneous density in the (surviving clusters).
This fact is however very nontrivial from the mathematical
viewpoint and the problem \(stefan), \(vel2)
may become ill--posed for some finite $\tau_2$.
It is however easy to verify that if the density
in the clusters is  that of one of the homegeneous phases,
then $V_2\equiv 0$ and so there is no macroscopic motion of the
interfaces on the
time scale $q=2$, but there may be a motion
on a longer time scale, i.e. if $q$ is larger.

{\bf The sharp interface II: the Mullins--Sekerka
symmetric solidification model ($q=3$).}
Let us assume that the initial condition satisfies
the extra condition
$\rhe_0(r)= \rho ^\pm +O(\epsilon)$ for every
$r$ away from the interface
and that, like for the case $q=2$,
 the density profile at the interface
is given by the rescaled instantonic solution
in the direction perpendicular to the interface.
It turns out that,
for such an initial condition, the relevant time scale
is $q=3$. Under the assumption of the stability
of the interface (see the discussion for
the $q=2$ case), we can describe the time evolution
of the density in terms of the evolution of the
interface, since the density in the cluster
is just the equilibrium one. The problem is then reduced
to find the geometrical motion of the interface
$\Gamma _{\tau_3}$. The analog of the
natural domain decomposition \(natdecomp) clearly
holds also in this case.
In the Appendix it is shown
 that for this model the surface tension
(i.e. the free energy density associated with the instantonic
solution $U$, see [\rcite{Butta}]) is
$$
S
= {1\over 2}\int _{\real \times \real}
 (z^\prime-z)
\partial _z U  (z) U (z^\prime)
\tilde J( z-z^\prime) \text{d} z \text{d} z^\prime
\(surftens)
$$
Let us furthermore denote by $K(r)$, $r$ on the interface, the sum
of the principal curvatures of the interface at $r$, that is $d-1$
times the mean curvature at $r$.

\proclaim{Statement 2}
In the setting explained above and at the level of
formal asymptotics, the motion of the interface $\Gamma_{\tau_3}$ is
generated by the velocity field $V_3(r)\nu(r)$, for $r$ belonging
to the interface $\Gamma_{\tau_3}$. The speed is given by the following
expression
$$
V_3(r)={ \sigma^\pm \over 2 \rho^+-1}
[\nu \cdot \nabla \mu_1]_-^+(r)
\(vel3)
$$
where $[\nu \cdot\nabla\mu_1]_-^+(r)$ denotes the jump of
the component normal to the interface of the function
$\mu_1$, which is continuous on $T^d$, at least twice
differentiable on $\Omega^\pm_{\tau_3}$ and solves
the static problem
$$
\cases
\Delta \mu _1 (r) =0 &\text{ for } r \in \Omega_{\tau_3} ^+
 \cup \Omega_{\tau _3}^- \\
\mu _1(r)= -{S K(r)/ (2\rho^+ -1)}
 &\text{ for } r \in \Gamma_{\tau _3}
\endcases
\(staticMS)
$$
\endproclaim

{\it Remark 1:} The global existence for the Mullins--Sekerka
evolution law presents severe mathematical difficulties.
We refer to [\rcite{Chen}] for these aspects of the problem.

{\it Remark 2:}
 The function $\mu_1$, which appears in \(vel3) and
\(staticMS), may be thought of just as an auxiliary
tool for computing the interface velocity.
It does however have a direct physical meaning:
it is the $O(\epsilon)$ correction to the chemical potential
away from the interface,
i.e. $\mu= \mu(\rho^\pm)+\epsilon \mu_1 +o(\epsilon)$
in the interior of the clusters.

{\bf Some remarks on the limit of vanishing temperature.}
Here we restrict ourselves to   the model defined by \(entropy) and
\(sigma). To stress the dependence on the parameter $\beta$
we add a subscript $\beta$ to $U$, $\sigma ^\pm$,
$\rho^\pm$ and $S$. In this case,
a natural question is what happens to the velocity $V_3$
in the Mullins--Sekerka model as $\beta $ goes to infinity.
In this limit the instantonic solution
$U_\beta$ approaches the Heaviside function (easily shown using
formula \(functional)),
 i.e. the characteristic
function of the positive semiaxis. Therefore, using
\(surftens),  we easily see that
$$
\lim_{\beta \rightarrow \infty}S_\beta= S_\infty\equiv
{1\over2}\int_0^\infty r \tilde J (r) dr
$$
and $S_\infty$ is  strictly positive and finite.
By \(staticMS)
the gradient of $\mu_1$ is uniformly bounded
 as long as the interface is smooth. So what really matters
to our analysis is the behavior, as $\beta $ becomes larger and larger,
of the factor $\beta \sigma^\pm_\beta /(2\rho_\beta ^+-1)$ in
\(vel3). While the denominator tends to one, the mobility
in the phase $\sigma _\beta$ vanishes exponentially
fast.
Our conclusion is then
$$
\lim_{\beta \rightarrow \infty}
V_3 =0
$$
and the limit is approached exponentially fast. So the
Mullins--Sekerka motion is rapidly frozen when the
temperature is lowered. A motion on the
time scale $q=4$ is expected to arise (surface diffusion), but this will
not be considered here.

\head
4. Existence, uniqueness and regularity
\endhead
\resetall
\sectno=4

We follow closely the notation of [\rcite{DL3}], to which we refer
for all the basic notions and results. In the first part
of this Section we work in $T^d$ (unit torus),
so that $L^2=L^2(T^d), \, H^1=H^1(T^d), \ldots$. $\tau>0$ is  fixed
and our basic (Hilbert) space is
$$
W\equiv W(0,\tau; \, H^1, H^{-1})\equiv
\left\{w: w\in L^2(0,\tau;H^1), \,
{{\text d} w\over {\text d}t} \in
L^2(0,\tau; H^{-1})
\right\}
\(spaceW)
$$
where $H^{-1}=(H^1)^\prime$, the dual of $H^1$ with respect to
the $L^2$ scalar product $(\cdot, \cdot)$.
The norm in $H^1$ will be denoted by
$\Vert \cdot \Vert$, the norm in $L^2$  by
$\vert \cdot \vert _2$ and the norm in $H^{-1}$  by
$\Vert \cdot \Vert_*$. In particular we have
$$
\Vert v \Vert_*= \sup _{u \in H^1}
{(u,v)\over \Vert u \Vert} =
 \vert (1-\Delta)^{-1/2} v)\vert_2
=\left( \sum_{k\in 2 \pi \Z^d}
{\left\vert \hat{v} (k) \right\vert^2 \over 1+k^2 } \right)^{1/2}
\(Hminus1norm)
$$
in which $v\in H^{-1}$ and $\hat{v}  $ is the Fourier transform
of $v$.
The space $W$ is equipped with the (Hilbert) norm
$$
\align
\Vert w \Vert_W&=
\left(
\Vert w \Vert_{L^2(0,\tau;H^1)}^2+
\Vert w^\prime \Vert_{L^2(0,\tau;H^{-1})}^2\right)^{1/2}
\\
&=
\left(
\int_0^\tau \left[
\Vert w(t) \Vert^2 +
\Vert w^\prime (t) \Vert _*^2\right]
\text{d}t \right)^{1/2}
\(Wnorm)
\endalign
$$
in which $w^\prime=
{\text d} w/ {\text d}t$.
We will often work with the following convex subset of $W$
$$
W_1\equiv \{ w\in W: 0\le w(t,x) \le 1 \
\text{for almost all } (t,x) \in [0,\tau]\times T^d \}.
\(W1)
$$

We say that $\rho \in W_1$ is a weak solution of \(maineq)
if for all $u \in H^1$
$$
{\text{d} \over \text{d}t}
\left(u, \rho\right)+\left(\nabla u, D(\rho) \nabla \rho- \sigma(\rho)
(\nabla J*\rho)\right)=0
\(mainrep)
$$
in the sense of ${\Cal D}^\prime((0,\tau))$ (${\Cal D}((0,\tau))$ is the
space of $C^\infty$ compactly supported real functions over $(0,\tau)$)
and
$$
\rho(0)=\overline{\rho}.
\(initweak)
$$
We recall that \(initweak) makes sense because
$W$ can be continuously imbedded in $C^0 (0,\tau;L^2)$, Th.1,
Ch.XVIII of [\rcite{DL3}],
and that $\text{d}
\left(u, \rho\right)/\text{d} t= (u, \rho^\prime)$.

The theorem we are going to state below
 holds under the assumptions made above and
under the following assumptions on $D$, $\sigma$ and $J$.
$$
D\in C^{0}([0,1])
\(assumeD)
$$
($u\in C^{k-}$, $k \in {\Z^+}$, means that the
derivatives of $u$
 of order $(k-1)^{th}$
are Lipschitz)
and that $1/c \le D \le c$ for some $c\ge 1$. Moreover
$$
\sigma\in C^{1-}([0,1]) \ \ \ \ \sigma (\rho )\ge 0 \ \text{ for all }
\rho \in [0,1], \ \ \ \ \sigma(0)=\sigma(1)=0.
\(assume3)
$$
Finally
$$
J\in C^2 (T^d), \ \ \ \ \ J(r)=J(-r) \text{ for all } r \in T^d.
\(assume2)
$$
\vskip 0.3 cm

\proclaim{Theorem 4.1}
There exists a unique solution to the problem
\(mainrep), \(initweak).
\endproclaim

\vskip 0.3 cm
{\it Proof.}
\vskip 0.1 cm

\noindent
{\it Existence.}
We will make use of the following
result.
\vskip 0.2 cm
\proclaim{Proposition 4.1}
For every $g \in L^2(0,\tau;L^2)$ the problem
$$
\cases
u^\prime=\nabla \{
D(u) \nabla u-\sigma(u) (\nabla J*g)\}\\
 \\
u(0)=\overline{\rho}
\endcases
\(intermediate)
$$
has a unique weak solution (in the same sense as \(mainrep)
and \(initweak)).
\endproclaim
\vskip 0.2 cm
{\it Proof of Proposition 4.1:} \(intermediate)
is a  parabolic problem for which existence in $W$
can be established
via fixed point theorems, see e.g. [\rcite{Lions}], IV.4,
(extending the definitions of $\sigma$ and $D$ outside
$[0,1]$ in a arbitrary (nice) way). The comparison
principle and the fact that $\sigma(0)=\sigma(1)=0$, so that
$u\equiv1$ and $u \equiv 0$ are solutions of \(intermediate)
with proper initial condition, immediately allows to obtain a solution
in $W_1$. Uniqueness is also standard and can be, for example,
established along the same line that we are going to
use below to prove uniqueness
for the original problem \(mainrep),\(initweak).
\qed

\vskip 0.2 cm

Proposition 4.1 defines a map $X:L^2(0,\tau;L^2)
\rightarrow L^2(0,\tau;L^2)$
$$
X(g)=u
\(map)
$$
and we notice that, since $0\le u\le 1$ almost everywhere,
$\Vert X(g)\Vert^2_{L^2(0,\tau;L^2)} \le \tau$. Directly from \(intermediate)
one has that
$$
(u(t), u^\prime (t))+
\int_{T^d} \vert \nabla u(t) \vert^2 D(u(t))=
\int_{T^d} \sigma (u(t)) (\nabla J*g(t)) \nabla u(t)
$$
so that by using the Cauchy--Schwarz and Young's inequalities, \(nondegen)
and the fact that $\sigma$ is bounded,
one obtains that there is a finite constant $c_1$ such that
$$
{1\over 2 }
{\text{d} \over \text{d} t}
\vert u(t) \vert _2^2 +{1\over 2c}
\vert \nabla u(t) \vert ^2_2 \le {c_1 \over 2} \vert g(t) \vert_2^2
\(mainbound)
$$
in which $c$ is the constant in \(nondegen). From \(mainbound),
by integrating
in time, we extract the following bound
$$
\int_0^\tau \Vert u(t) \Vert ^2 \text{d}t
\le c_2 \left[ \Vert g \Vert ^2_{L^2(0,\tau;L^2)}+1\right]
\(bound1)
$$
for some finite constant $c_2$ (independent of the initial condition).
Moreover
$$
\Vert u^\prime \Vert_*=
\sup_{u_1 \in H^1: \Vert u_1 \Vert =1}
\left (
\int_{T^d} \left[
\sigma (u) (\nabla J*g) \nabla u_1- D(u) \nabla u \nabla u_1\right]
\right)
$$
which by  the Cauchy--Schwarz inequality implies that there is
 a $c_3$ such that
$$
\int_0^\tau \Vert u^\prime (t) \Vert ^2_* \text{d}t
\le c_3
\left[ \Vert g \Vert ^2_{L^2(0,\tau;L^2)}+1\right]
\(bound2)
$$
so that by \(bound1) and \(bound2)
we have that there exists $c_4$ such that
$$
\Vert u \Vert_W\le c_4 \left[ \Vert g \Vert ^2_{L^2(0,\tau;L^2)}+1\right]
\(bound3)
$$
which implies
 that $X$ is compact, since $W$ is compactly imbedded in $L^2(0,\tau;L^2)$,
see Proposition 4.2, Section IV of [\rcite{Lions}].
$X$ is also continuous since
given a sequence $\{g_n\}$ which converges in $L^2(0,\tau;L^2)$ to
$g$, the sequence $\{ X(g_n)\}=\{u_n\}$ is relatively compact
in $L^2(0,\tau;L^2)$ and weakly relatively compact in W
(by \(bound3)). This in particular implies that
$D(u_n)\nabla u_n$ converges weakly along subsequences
 in $L^2(0,\tau;L^2)$ to
$D(u)\nabla u$, since $D(\cdot)$ is $C^0$ and bounded.
Hence any limit point $u$ of $\{ u_n\}$ (weakly in $W$ and
strongly in $L^2(0,\tau;L^2)$)  solves \(intermediate), which has a unique
solution, and the continuity of $X$ follows.

Summing  up, we have  obtained that the continuous map
$X$ takes the
closed convex set
$$
\{u\in L^2(0,\tau;L^2):\, \Vert u \Vert_{L^2(0,\tau;L^2)}^2 \le  \tau \}
$$
into itself and the image of this set under $X$
is relatively compact (in particular $X(u)\in W_1$ and
$\Vert u \Vert_W \le c_4 [1+\tau^{1/2}]$).
 By the Schauder fixed point Theorem
([\rcite{GT2nd}], Corollary 11.2) X has a fixed point
$\rho \in W_1$ which satisfies \(mainrep), \(initweak).
This completes the proof of the existence.

\vskip 0.2 cm
\noindent
{\it Uniqueness.}

Call $\rho_1$ and $\rho_2$ two solutions of \(mainrep)
such that $\rho_1(0)=\rho_2(0)$. Set $\lambda (\rho)=
\int_0^\rho D(\rho^\prime) \text{d}\rho^\prime$, so that
$\lambda\in C^{1}$ and $\lambda^\prime(\rho)\ge 1/c>0$.
We have
$$
{1\over 2}
{\text{d} \over \text{d}t}
\Vert \rho_1 (t)-\rho_2(t) \Vert^2_*=
-\left(\nabla (\rho_1(t)-\rho_2(t)),
(1-\Delta )^{-1}
\nabla (\lambda (\rho_1(t))-\lambda (\rho_2(t)))\right)+
$$
$$
\left(\nabla (\rho_1 (t)-\rho_2(t)), (1-\Delta)^{-1} \nabla(
\sigma(\rho_1) \nabla J* \rho_1(t)-
\sigma(\rho_2) \nabla J* \rho_2(t))\right).
$$
We then have
$$
\align
{1\over 2}
{\text{d} \over \text{d}t}
\Vert \rho_1 (t)-\rho_2(t) \Vert^2_*=&
-\left(\rho_1(t)-\rho_2(t),
(\lambda (\rho_1(t))-\lambda(\rho_2(t)))\right)\\
&
+\left(\rho_1(t)-\rho_2(t),
(1-\Delta )^{-1}
(\lambda (\rho_1(t))-\lambda (\rho_2(t)))\right)
\endalign
$$
$$
+
\left(\nabla (\rho_1 (t)-\rho_2(t)), (1-\Delta)^{-1} \nabla(
\sigma(\rho_1) \nabla J* \rho_1(t)-
\sigma(\rho_2) \nabla J* \rho_2(t))\right).
$$
Since $\lambda^{\prime }\ge 1/c>0$ and by the Young and Cauchy--Schwarz
inequalities, we obtain that for any $\epsilon_1, \epsilon_2 >0$
$$
{1\over 2}
{\text{d} \over \text{d}t}
\Vert \rho_1 (t)-\rho_2(t) \Vert^2_*\le
-{1\over c}\left\vert\rho_1(t)-\rho_2(t)\right\vert_2^2+
{\epsilon_1^2 \over 2}
\Vert \lambda (\rho_1) -\lambda (\rho_2)\Vert_*^2
+{1\over 2 \epsilon _1^2}
\Vert \rho_1 -\rho_2\Vert_*^2
$$
$$
+{1\over2 \epsilon_2^2}
\Vert \nabla (\rho_1 - \rho_2)\Vert_*^2+
{\epsilon_2^2 \over 2}
\Vert \nabla [\sigma(\rho_1) \nabla J* \rho_1(t)-
\sigma(\rho_2) \nabla J* \rho_2(t)]\Vert _* .
\(Schwarz1)
$$
Since for all $v \in L^2$, $\Vert v\Vert_* \le \vert v \vert_2$,
for all $u \in H^1$, $\Vert \nabla u \Vert_* \le \vert u \vert _2$, and
$$
\left\Vert \nabla [\sigma(\rho_1) \nabla J* \rho_1(t)-
\sigma(\rho_2) \nabla J* \rho_2(t)]\right\Vert _*
\le
\left \vert [\sigma(\rho_1) \nabla J* \rho_1(t)-
\sigma(\rho_2) \nabla J* \rho_2(t)\right\vert_2
$$
$$
\le \big\vert [\sigma(\rho_1) -\sigma(\rho_2) ]
\vert \nabla J * \rho_1 \vert \big\vert_2 +
\big\vert \sigma(\rho_2)
\vert \nabla J *
(\rho_2-\rho_1)
\vert \big\vert_2
\le c_5 \vert \rho_1 -\rho_2 \vert_2
$$
where $c_5$ depends only on the Lipschitz constant of $\sigma$ and on $\vert
\nabla J \vert _2$.  From \(Schwarz1), choosing $\epsilon_1$ and $\epsilon_2$
small, we obtain that there is $c_6$ such that
$$
{1\over 2}
{\text{d} \over \text{d}t}
\Vert \rho_1 (t)-\rho_2(t) \Vert^2_*\le c_6 \,
\Vert \rho_1 (t)-\rho_2(t) \Vert^2_*
$$
which, since $\Vert \rho_1 (0)-\rho_2(0) \Vert_*=0$,
implies that $\Vert \rho_1 (t)-\rho_2(t) \Vert_*=0$ for
all $t \in [0,\tau]$  and,
since $\rho_1, \rho_2\in W$, $\Vert \rho_1 (t)-\rho_2(t) \Vert=0$ which
implies the result.
\qed
\vskip 0.2 cm

\noindent
{\bf Remark 4.1:}
We observe that since $\rho\in C^0(0,\tau;L^2)$,
$\nabla J *\rho\in C^0(0,\tau;C^1)$ (since $J\in C^2$).
Once we have a solution of \(mainrep), \(initweak),
we have also a solution of
\(intermediate)
for $g=J *\rho$. It is then a standard result (see e.g. [\rcite{Lady}])
that, if $\overline {\rho}\in C^2$ and if $D$ and $\sigma $ are $C^1$,
 \(intermediate) has a classical solution, which is unique
in the class of functions considered in Theorem 4.1 and Proposition 4.1.
Therefore under
the same assuptions one has a classical solution for our
original problem \(mainrep), \(initweak).
 One can obtain further regularity properties
by making further assumptions on $D$, $\sigma$
and $J$.

\vskip 0.3 cm

We now establish the existence and uniqueness (in a proper sense)
 of the istantonic
solution and we give some of its properties. Sometimes
in what follows we keep
the notation with partial derivatives also for functions
of one variable, for uniformity.

Consider the one dimensional setting with $a=\infty$, i.e.
we are working on $\R$.
Assume 
\(extra1),   \(assume4), \(GHS) , 
that $f_c$ is non convex
and all the 
assumptions made for Theorem 1.
For consistency with the sharp asymptotic
computations, below
we use $\tilde{J}$ for $J$ in the one dimensional setting
(recall \(Jtilda)). Take $\tilde{J}$ to be compactly supported.

\proclaim{Proposition 4.2}
Under the assumptions stated above, there exists
a function $U\in C^1 (\real)$ such that for all $v \in C_0^\infty(\real)$
$$
\int_\real \partial_z v(z)
\left[ D(U) \partial _z U(z) - \sigma (U(z))\partial_z (\tilde{J} *U)(z)
\right] \text{d}z =0.
\(deq1stat)
$$
$U$ satisfies:\hfill\break
(1) $U-1/2$ is odd;
\hfill\break
(2) $U^\prime (z)>0$ for all $z \in \real$;
\hfill\break
(3) There exists $m>0$ such that
$$
\lim_{z \rightarrow \pm \infty} e^{m\vert z \vert}
\left\vert U(z)-\rho^\pm \right\vert=0.
\(exptail)
$$
Moreover if a non constant non decreasing function
$u\in H^1_{loc}(\real)$,
 $u(z) \in (0,1)$ for all $z\in \real$,
satisfies \(deq1stat) with $U=u$ for all $v \in C_0^\infty(\real)$, then
there exists $\overline{z}\in \real$ such that
$u(z)= U(z-\overline{z})$.
\endproclaim

\vskip 0.3 cm
{\it Proof:}
take $u$ as in the statement of the Proposition above.
Then there exists $C \in \real$ such that
$$
D(u)\partial _z u(z)
-\sigma (u(z)) (\tilde{J}^{\prime} *u)(z)
=C
\(isCC2)
$$
for almost all $z$.
Since the second term in the left--hand side
vanishes as $z \rightarrow \infty$ and 
 $D>1/c$, then $C\ge 0$
and
$\partial_z u(z)\ge cC/2$ for $z$ sufficiently large, so that $C=0$.
Since $u(z) \in (0,1)$ for all $z$,
$\vert f^{\prime \prime}\vert$ (recall \(nondegen))
is bounded on every compact subset of $\real$ and $\sigma(u)>0$.
Hence we have
$$
\partial _z
\left[  f^\prime (u(z)) -  (\tilde{J} *u)(z)
\right]=0
\(iszero)
$$
which, using the fact that $f^{\prime \prime}>0$, is equivalent
to
$$
u(z)=(f^\prime)^{-1}\left(  (\tilde{J} *(u-1/2))(z)
+h\right)
\(integraleq)
$$
for some $h\in \real$.
In [\rcite{DMDP}] and [\rcite{DOPTrend}] it is shown that,
in the class of functions
we are considering and for the particular case
$f^\prime (\rho)=\log (\rho/1-\rho)$, so that
$(f^\prime)^{-1}(x)=
(1/2)[\tanh(x/2)+1]$ there is a solution to \(integraleq)
in the case $h=0$ and [\rcite{DOPTrend}] it is unique in a 
class larger than
the one we consider. Their approach
can be repeated in our context, but 
the general result in [\rcite{Chen2}], in particular
Remark 5.2 $(ii)$,
keeping in mind that we have the symmetry condition \(extra3),
covers the case considered here, establishing thus existence
and uniqueness for $h=0$: property (1) follows from
\(extra3) as well.
Property (2) is established by  the argument on page 706
of [\rcite{DOPTrend}] and
 property (3) is established by repeating the proof of
   Proposition 2.2 in [\rcite{DOPTedimb}]
for the general $f$ considered here. We have thus established the existence
of the function $U$ claimed in  the Proposition and
that $U$ solves \(functional).
We remark that in
[\rcite{DGP},\rcite{DOPTrend},\rcite{DOPTedimb}],
 $\hat{J}(0)=1$ and $J$ is multiplied by $\beta>0$,
while
in our case the role of $\beta $ is played by $\hat{J}(0)$,
and the density profile is given in terms of the
magnetization variable $m=2 \rho-1$.

We are left with showing that there is no solution to
\(integraleq) for $h \not=0$.
In [\rcite{Chen2}] (see also [\rcite{DGP}])
it is proven that the evolution equation
$$
\partial_t u(r,t)=-u(r,t) + (f^\prime)^{-1}\left(  (\tilde{J}
*(u(\cdot,t)(r)-1/2) +h\right)
\(Chen2)
$$
has a unique solution in the class of travelling wave solutions
if $f^\prime (\rho)-\hat{J}(0)(\rho-1/2)-h=0$ has three distinct solutions
and this solution has
 non zero speed for $h \not=0$. In particular this implies that there is no
solution with zero speed, i.e. stationary, in the case of three
distinct roots. Due to \(GHS), $f^\prime (\rho)-\hat{J}(0)(\rho-1/2)-h$
can have at most three roots. If it has only one root $\rho_1$, then
$u$ must be constant, since the asymptotic values of $u$ must
be equal to $\rho_1$. We are then left with the
case in which
 there is a root $\rho_1$ and a double root $\rho_2$.
With no loss of generality let us assume that $\rho_2>\rho_1$ so that
$\lim_{z \rightarrow \infty} u(z)=\rho_2$.
Due to \(GHS) we have that
$$
f^\prime(u)-\hat{J}(0)(u-1/ 2)-h=
\tilde{J}
*(u-1/2) -\hat{J}(0) (u-1/ 2)\ge 0
\(onlytwo)
$$
whenever $u(z)\in [\overline{u}, \rho_2]$, for some
$\overline{u}< \rho_2$, and the inequality is strict 
for $u(z)\in [\overline{u}, \rho_2)$.
If $z_0=\inf\{z: u(z)<\rho_2\}<\infty$, then a contraddiction 
is easily established by \(onlytwo) with $z=z_0$.
If $z_0=\infty$, since $u$ is monotonic, \(onlytwo) 
is valid for all $z$ sufficiently large. However
this implies that $u$ is unbounded, which
establishes the desired contraddiction and this
completes the proof of uniqueness.
 We remark that uniqueness
holds with no need of  \(GHS)
 if we restrict ourselves  to establish uniqueness in the class of functions
which have asymptotic at $z=\pm \infty$ which are symmetric
with respect to $1/2$, since this imposes $h=0$ in \(integraleq).
\qed
\vskip 0.3 cm

We just remark that the regularity of the istantonic
solution can be easily read out  from \(integraleq) in terms
of the regularity of $J$ and $f$.

\head
5. Formal asymptotic expansions
\endhead
\resetall
\sectno=5

The computations regarding Statement 1 and Statement 2
follow closely the scheme of R. L. Pego's work
[\rcite{Pego}].
We start by expanding the density profile in the bulk
({\sl outer expansion}) and then we will do the same
close to the interface ({\sl inner expansion}): we will
finally match the values at the boundaries.

We recall that \(extra1), \(extra2) and \(extra3) are assumed.
 With abuse of notation,  $J(R)$ ($R\in \real ^+$)
 will mean $J(r)$, for $\vert r \vert=R$.
Until further notice, $\tau=\tau_q$, any value of $q$ (see \(rhoeps)
and \(mainresc)).

{\it The outer expansion}:
we write
$$
\rho^\epsilon (r,\tau )=\rho_0 (r,\tau)+
\eps \rho_1 (r, \tau) +\eps ^2 \rho_2 (r,\tau)+\ldots
\(bulkrhoG)
$$
$$
\sgm^\epsilon (r,\tau )\equiv
\sgm(\rho^\epsilon (r, \tau))=
\sgm_0 (r,\tau)+
\eps \sgm_1 (r, \tau) +\eps ^2 \sgm_2 (r,\tau)+\ldots
\(bulksgmG)
$$
$$
\mu^\epsilon (r,\tau )=\mu_0 (r,\tau)+
\eps \mu_1 (r, \tau) +\eps^2 \mu_2 (r,\tau)+\ldots
\(bulkmuG)
$$
By Taylor expansion, using  \(mueps), one  obtains directly
$$
\sigma _0(r,\tau)=\sigma(\rho_0(r,\tau))
\(bulksgm0)
$$
$$
\sgm_1(r,\tau)=\rho_1 \sigma^\prime(\rho_0)
\(bulksgm1)
$$
$$
\sgm_2=\rho_2 \sigma^\prime(\rho_0)+\rho_1^2 {
\sigma^{\prime\prime}(\rho_0)\over 2}
\(bulksgm2)
$$
$$
\mu_0=f^\prime (\rho_0)
-{\hat J}(0) \rho_0
\(bulkmu0)
$$
$$
\mu_1 =f^{\prime \prime} (\rho_0) \rho_1 -
{\hat J}(0) \rho_1
\(bulkmu1)
$$
$$
\mu_2= f^{\prime \prime\prime}(\rho_0) \rho_2
+f^{\prime \prime} {\rho^2_1 \over 2}
 -
{\hat J}(0) \rho_2 -
{{\bar J}\over 2}\Delta \rho_0
\(bulkmu2)
$$
in which
${\bar J}= (1/d)\int J(r) r ^2
\text{d} r$ and we recall that
$\hat J (0)=\int J(r) \text{d} r$
\vskip 0.6 cm

{\it The inner expansion}: We want to expand the density
profile $\rhe$ in a small neighborhood of
the interface $\Gamma _\tau$, $\tau \ge 0$.
One possible choice of the neighborhood is
$$
{\Cal N}_\epsilon =\{r\in \reald :
\vert \phi(r, \Gamma _\tau)\vert \le \epsilon ^a\}
\(tub)
$$
where $a\in (0,1)$ (recall that
$\phi$ denotes the signed distance).
let us set $z=\phi(r, \Gamma_\tau) /\epsilon$.
If $\eps $ is sufficiently small (depending on
the minimum  radius of curvature of the interface) then
for each $r\in {\Cal N}_\epsilon$ there is
a unique $x=x(r)\in \Gamma_\tau$ such that $r=x + \nu(x) \eps z$.
Notice that
$$
\nu(x(r)) =\nabla \phi (r, \Gamma_\tau)
\(nu)
$$
for any $r\in {\Cal N}_\epsilon$. Moreover we define
$$
V(x)\equiv V(x(r))={\partial \phi\over \partial \tau}(r,\tau)
\(V)
$$
Note that $\partial \phi /\partial \tau$ is independent of
$r\in {\Cal N}_\epsilon$, as long as $x(r)$ is held fixed.
For $r \in {\Cal N}_\eps$
 will look for expansions of the form
$$
\rho^\epsilon (r,\tau )=\trho_0 (z,r,\tau)+
\eps \trho_1 (z, r, \tau) +\eps ^2 \trho_2 (z,r,\tau)+\ldots
\(bulkrhoG)
$$
$$
\sgm^\epsilon (r,\tau )=\tsgm_0 (z,r,\tau)+
\eps \tsgm_1 (z,r, \tau) +\eps ^2 \tsgm_2 (z,r,\tau)+\ldots
\(bulksgmG)
$$
$$
\mu^\epsilon (r,\tau )=\tmu_0 (z,r,\tau)+
\eps \tmu_1 (z,r, \tau) +\eps ^2\tmu_2 (z,r,\tau)+\ldots
\(bulkmuG)
$$
in which any quantity $w(z,r,\tau)$ is defined
for every $r$ in ${\Cal N}_\eps$ and  $w(z,r+c\nu,\tau)=
w(z,r,\tau)$,
$\nu =\nu(x(r))$,
 for all $c\in \real$  such that $r+c\nu \in {\Cal N}_\eps$,
so
 that $\nu  \cdot \nabla w (z,r,\tau)=0$.
The following formulas will  be useful
$$
\nabla \rho(r, \tau)= \eps^{-1} \nu(r, \tau)\partial_z \trho (z, r, \tau)
+\nabla_r \trho (z,r, \tau)
\(newgrad)
$$
$$
\partial_\tau \rho(r, \tau)=
\partial_\tau \trho (z, r, \tau)+
\eps ^{-1} V(r, \tau) \partial_z \trho (z, r, \tau)
\(newdt)
$$
The inner expansion for $\sigma$ is identical to the outer one
$$
\tsgm _0= \sigma (\trho _0)
\(insgm0)
$$
$$
\tsgm_1= \sigma^\prime(\trho_0) \trho _1
\(insgm1)
$$
$$
\tsgm_2=
\sigma ^{\prime \prime} (\trho_0) \trho_2 +
\sigma ^{\prime } (\trho _0) {\trho_1 ^2 \over 2}
\(insgm2)
$$
More complicated is the computation
of $\tmu _i$: for the moment we give it only to first order
$$
\tmu_0= f^\prime (\trho _0)
- \tJ \star \trho _0
\(inmu0)
$$
in which $\star$ stands for the convolution in the $z$ variable,
that is
$$
(\tJ \star w) (z, r, \tau)=
\int_\real \tJ (z-z^\prime) w(z^\prime, x, \tau)
\text{d} z^\prime
$$
and we recall that $\tJ$ is defined in \(Jtilda).

The relation between $\Gamma_\tau$  and $\trho_i$ is given by the following
{\sl front centering condition} [\rcite{Pego}]:
$$
\int_{-\infty}^{+\infty}
U ^\prime (z)
\left[\trho (z,r,\tau) -U (z) \right]
\text{d} z=0
\(centering)
$$
where $\trho=\sum_i \epsilon^i \trho_i $ and the coefficient
of each power of $\epsilon$ in \(centering) is required to vanish.

\vskip 0.7 cm
{\it Matching conditions:}
We will match the outer and the inner expansion
of the chemical potential  order by order
by requiring formally that
$$
(\mu_0+ \eps \mu_1 +\eps ^2 \mu_2 +\ldots )
(r+\eps z\nu(r, \tau), \tau)
\approx (\tmu _0+\eps \tmu_1 +\eps ^2 \tmu_2 +\ldots )
(z,r,t)
\(match)
$$
with $z=\epsilon ^{1-a}$, $a \in (0,1)$. By expanding
the left--hand side of \(match) we obtain (see
[\rcite{Pego}])
$$
\mu_0^{\pm} (r,\tau)=\lim_{z \rightarrow \pm \infty}
\tmu_0 (z,r,\tau)
\(match0)
$$
$$
(\mu_1^\pm +z \nu\cdot \nabla \mu_0^\pm )(r,\tau)=
\tmu _1(z,r,\tau) +o(1) \ \ \ \text{as } z \goes \pm \infty
\(match1)
$$
$$
\left(\mu_2^\pm +z\nu\cdot\nabla\mu_1 ^\pm +
{1\over 2} z^2 (\nu \cdot \nabla)^2 \mu_0^\pm
\right)(r,\tau)=
\tmu _2 (z,x,\tau) +o(1)
\ \ \ \text{as } z \goes \pm \infty
\(match2)
$$
in which $\mu_i^\pm   (r, \tau) = \lim_{t \goes 0^\pm}
\mu_i (r+\nu t, \tau)$, and analogous formula for the derivatives.

\vskip 0.7 cm
{\it Verification of Statement 1: the case $q=2$}. The starting
point is equation \(mainresc) and we set $\tau= \tau_2$.
By using
\(nu), \(newgrad) and  \(newdt),
 we obtain that the highest
order arising in the inner expansion of \(mainresc)
($O(\epsilon^{-2})$)
 yields the equation
$$
\partial _z\left(
\tsgm_0 \partial_z \tmu_0\right)=0.
\(setinst)
$$
Using now the explicit expressions for $\tsgm_0$ \(insgm0) and
$\tmu_0$ \(inmu0) we see that the solution
$\trho_0$ of \(setinst) coincides with the stationary
solution of \(main1d), that is
$$
\trho_0 (z,r,\tau)=U(z)
\(set0)
$$
Moreover by using \(functional) we obtain
$$
\tmu_0 (z,r,\tau)={\hat{J}(0) \over 2}
\(implied0)
$$
The leading order of \(mainresc) in the outer expansion ($O(1)$)
is
$$
\partial_\tau \rho_0 =
\nabla \left(
\sigma_0 \nabla \mu_0 \right)
\(nlPDE)
$$
which by the assumptions on \(bulksgm0),\(bulkmu0) (see \(extra3))
and on the initial condition
is a nondegenerate parabolic PDE. \(nlPDE) has to be
supplemented with the boundary conditions coming
from the matching rule \(match0) and the asymptotic value
of $\trho_0$, given by \(asymptph). This completes the derivation
of  \(stefan) and we are left with the identification
of the speed of the interface. To do this we have to look at
the next order in the inner expansion
($O(\eps^{-1})$). From \(V), \(newgrad), \(newdt) and the fact that
$\tmu_0$ is constant \(implied0) we obtain
$$
V\partial_z \trho _0=
\partial_z
\left(
\tsgm _0 \partial_0 \tmu_1 \right)
\(intvel2)
$$
Recall that $\trho_0= U$ and integrate in $z \in \real$
to obtain
$$
V_2\equiv V={\sgm ^\pm \left[
\partial _z \tmu_1 \right] _{-\infty }^{+\infty}
\over \left[ U \right] _{-\infty} ^{+\infty}}
\(velocity2)
$$
in which $[w]_{-\infty} ^{+\infty}= w(+\infty, r,\tau)-w(-\infty, r, \tau)$
and recall that $\sgm (\rho  ^+)=\sgm (\rho  ^-)
\equiv \sgm ^\pm$. Finally the matching condition
\(match1) implies
$$
\left[
\partial_z \tmu _1 \right]_{-\infty}^{+\infty}=
\left[
\nu \cdot \nabla \mu_0
\right]_-^+
\(vel2def)
$$
 and the notation
$[\cdot ]^+_-$ is introduced in
Statement 1.
Formula \(velocity2) together with \(vel2def) implies
\(vel2).
This completes the verification of Statement 1.
\qed

\vskip 0.7 cm
{\it Verification of Statement 2: The case $q=3$.}

In this section $\tau=\tau_3$.
By assumption $\rho_0(r,0)=\rho^\pm$
in $\Omega^\pm_0$ and the the highest order in the
outer expansion of \(mainresc) implies that
$\rho_0(r,\tau)=\rho^\pm$
in $\Omega^\pm_\tau$ also for $\tau >0$.
The lowest order in the inner expansion
is still $O(\eps ^{-2})$ and it yields once again \(setinst), so that
$\trho _0= U $, and \(implied0). The next order in the
outer expansion ($O(\eps)$) gives
$$
\partial _\tau \rho_0=
\nabla \left(
\sgm_0 \nabla \mu_1 +\sgm _1 \nabla \mu_0 \right)
\(nextout)
$$
but since
$\partial _\tau \rho_0=0$,
$\nabla \mu_0=0$ and $\sgm_0 (\rho ^\pm )=\sgm ^\pm
\not=0$ we obtain
$$
\Delta \mu_1 =0
\(MSbulk)
$$
Since $\partial_z \tmu_0 =0$
\(implied0), the $O(\eps ^{-1})$ in the inner
expansion gives
$$
\partial _z \left\{
\tsgm_0 \partial_z \tmu_1\right\}
=0
\(abcd)
$$
so that $\partial_z \tmu_1=0$,
since $\tsgm_0\not=0$ and
by \(match1) $\tmu _1 (z)$ approaches (as $z$ approaches
$\pm \infty$)
the value of $\nu \cdot \nabla \mu_0$ near the interface and
$\nabla \mu_0\equiv 0$. Hence
$$
\tmu _1 (z,r,\tau)=C(r,\tau).
\(mu1isc)
$$
 We will verify below that
$$
\tmu_1 (z, r,\tau)=
\trho_1 (z, r,\tau) f^{\prime \prime} (\trho_0 (z,r,\tau))
-
\tJ * \trho_1 -K(x(r))
\int \tJ (z-z^\prime)
\trho _0 (z^\prime , r ,\tau )
\text{d} z^\prime
\(inmu1)
$$
and putting together \(inmu1) and \(mu1isc) we obtain
$$
\trho _1 (z,r,\tau) f^{\prime \prime} (U (z))
 -
(\tilde{J}\star \trho_1)(z,r,\tau)-K(x(r))
\int_{\real}
(z-z^\prime) \tilde{J} (z-z^\prime)
\text{d} z^\prime
=C(r, \tau)
\(intermedia)
$$
Multiply both sides of \(intermedia)
by $U ^\prime (z)$ and integrate in the variable $z$  over all
$\real$.
Since $\int \trho_1 [\partial_z  f^\prime (U)/-
\tJ \star U ^\prime ] \text{d} z=0$, the expression
between square brackets being zero because it is simply
$\partial_z \tmu_0$,
we obtain
$$
-K(r) \int _{\real \times \real}
(z-z^\prime)
J(z-z^\prime) U (z^\prime) U ^\prime (z)
\text{d} z \text{d} z^\prime
=C(r, \tau) [U]_{-\infty}^{+\infty}
\(intermedia2)
$$
In the left--hand side of \(intermedia2) we recognize
the surface tension of the model \(surftens) and so
$$
\tmu_1=C=-{KS\over \left[ U \right]_{-\infty}^{+\infty}}
\(C)
$$
and in particular $\tmu_1$ is independent of time.
The velocity of the interface is once again retrieved
by looking at the next order in the inner expansion ($O(1)$)
$$
VU^\prime  = \partial _z \left(
\tsgm _0 \partial _z \tmu _2 \right)
\(nextordin)
$$
so that by integration
$$
V= {
\sigma ^\pm \left[ \partial_z \tmu _2 \right]_{-\infty}^{+\infty}
\over \left[ U \right]_{-\infty}^{+\infty}
}
\(fgh)
$$
and by \(match2) we finally obtain
$$
V_3\equiv V=
{\sigma ^\pm \left[ \nu\cdot \nabla \mu_1
\right]_{-}^{+} \over \left[ U \right]_{-\infty}^{+\infty}}
\(vel3der)
$$
that is \(vel3). On the other  hand \(match1) and \(C)
give us
the boundary conditions for $\mu_1$, which then solves
$\Delta \mu_1 =0$ with
$$
\mu_1 (r)=
-{ K(r) S \over \left[U \right]_{-\infty}^{+\infty}}
\(staticMSder)
$$
when $r\in \Gamma_\tau$, and the boundary value
problem \(staticMS) is derived.
We are left with showing that the first order correction
to the chemical potential in the bulk is given
by \(inmu1). The expression in the right--hand side
of \(inmu1) consists of three terms: the first one is simply
$f^{\prime \prime }(\trho _0) \trho_1$. The other two originate
from the $O(\eps)$ contributions  to the integral term
$$
\int
J_\eps (r-r^\prime)
\left[ \trho_0 \left(\phi(r^\prime , \Gamma_\tau)/\epsilon , r ^\prime,
\tau\right)
+ \eps \trho_1 \left(\phi(r^\prime , \Gamma_\tau)/\epsilon , r ^\prime,
\tau\right) +O(\eps ^2) \right]
\(start1)
$$
The $O(1)$ contribution to \(start1)
is simply $\tilde J \star \trho_0$ and if we subtract it from
\(start1) and we use $\trho_0=U$ we obtain
$$
\left[
\int J_\eps (r-r^\prime ) U (z^\prime) \text{d} r^\prime
-\left( \tilde J \star  U \right) (z)\right]
+\eps \tilde J \star \trho_1 +O(\eps ^2)
\(start2)
$$
in which $z$ (respectively $z^\prime$) is the rescaled distance of $r$
(respectively $r^\prime$) from the interface, as before,
i.e. $z=\phi (r, \Gamma _\tau)/\epsilon$ and
$z^\prime=\phi (r^\prime, \Gamma _\tau)/\epsilon$.
We are then left to show that the term between square
brackets is of order $\epsilon$ and that its value divided by
$\epsilon$  coincides to leading order  with the third term in the
right--hand side of \(inmu1). By choosing  an appropriate
frame of reference, we can assume that $r_i=0$ for $i=1,\ldots , d-1$
and $r_d \equiv z \eps$ and that in this coordinate system the upper half
plane $\{ r\in \real ^d: r_d =0\}$ is tangent to $\Gamma_\tau$.
We will also orient the $i$--axis, $i=1,\ldots, d-1$ along the directions
of principal curvatures. The curvature in the direction $i$ will be
denoted by $k_i$, so that $K=\sum_{i=1}^{d-1} k_i$.
It is moreover convenient to go back to the original coordinates,
that is to $x=r \eps ^{-1}=(0, \dots, 0, z)$ and
$x^\prime = r^\prime \eps ^{-1}$: $z^\prime =\phi (r^\prime , \Gamma_\tau )
/\eps$ is left unchanged, since $
\phi (x^\prime , \eps ^{-1} \Gamma_\tau )=\phi(\eps x^\prime, \Gamma_\tau)
/\eps$.
We have
$$
\int J_\eps (r-r^\prime)
U (z^\prime)
\text{d} r^\prime=
\int J(x-x^\prime) U(z^\prime) \text{d} x^\prime
\(changefc)
$$
Notice that since $J$ is supported in $B_1(0)$,the
$d$--dimensional
ball of radius $1$, centered at $0$, we are only interested in
$\vert x^\prime \vert =O(1)$
and $z^\prime=O(1)$.
We will parametrize $\epsilon ^{-1} \Gamma_\tau$ as a function $f_\epsilon :
\real ^d \rightarrow \real$ around the point $0 \in \real^{d-1}$,
that is $\{ r\in \real ^d: \vert(r_1,\ldots ,
r_{d-1} )\vert \}\cap \eps ^{-1} \Gamma _\tau =
\{ (q, f_\eps (q)): q\in \real ^{d-1}, \
\vert q \vert \le 1 \}$. By the fact that
$\Gamma_\tau$ is assumed to be smooth
we have
$$
f_\eps (q)=
-\sum _{i=1} ^{d-1}
{\epsilon k_i q_i^2 \over 2} +O(\eps^2)
\(expandgamma)
$$
for $\vert q \vert \le 1$.
Our aim is to write $z^\prime$ explicitely as a function
of $x^\prime$ so that  it will make the integral
in \(changefc) more explicit.
Since $x^\prime \in \eps ^{-1} {\Cal N}_\eps$ we have
$$
x^\prime =
(q, f_\eps (q)) +
{(-\nabla f_\eps ,1) \over
\sqrt {1+\vert \nabla f_\eps  \vert ^2 }}z^\prime +O(\eps ^2)
\(normalf)
$$
and, by using \(expandgamma), for $i=1, \ldots , d-1$
we obtain
$$
x_i^\prime =q_i +\eps k_i q_i z^\prime +O(\eps^2)
\(expandx1)
$$
and
$$
x_d ^\prime =z^\prime -\eps \sum _{i=1}^{d-1}
{k_i q_i ^2 \over 2} +O(\eps ^2)
\(expandxd)
$$
which imply
$$
z^\prime =x^\prime _d
+\eps \sum _{i=1}^{d-1}
{k_i {x_i^\prime} ^2 \over 2}
+O(\eps^2)
\(finallyexp)
$$
Let us now go back to \(start2).
The term in the square brackets is equal to
$$
\int J(x-x^\prime )
\left[ U (z^\prime) -U (x^\prime _d)\right]
\text{d} x^\prime =
 \eps
\int J(x-x^\prime )
\left[
U^\prime  (x^\prime_d)
\sum_{i=1} ^{d-1}
{k_i (x_i ^\prime)^2 \over 2}
\right]\text{d} x^\prime+O(\eps ^2)
\(qrtw)
$$
in which we used the definition of $\tilde J$ \(Jtilda)
and \(finallyexp). We are then left with  evaluating
$$
\int J(x-x^\prime )
\left(
\sum_i {k_i (x_i ^\prime)^2 \over 2}
\right)
U^\prime  (x^\prime _d )  \text{d} x^\prime
\(qrtz)
$$
in which $x=(0, \ldots , 0, z)$.
Set $q=(x^\prime _1, \ldots , x^\prime _{d-1})\in
\real ^{d-1}$, $x^\prime _d =z^\prime$ and
$R= \sqrt{q^2 +(z-z^\prime)^2}$.
Then the expression in \(qrtz)
is equal to
$$
{1\over 2} \sum _i k_i \int
J(R) U^\prime (z^\prime) q_i ^2 \text{d} q
\text {d} z^\prime
$$
$$
={1\over 2}
\left(\sum_{i=1}^{d-1} k_i \right)\int
J(R) U^\prime (z^\prime) q_1 ^2 \text{d} q
\text {d} z^\prime
$$
$$
=-{K\over 2}
\int
J^\prime(R)
\left({z^\prime -z \over R } \right)
U (z^\prime) q_1^2
\text{d} q
\text {d} z^\prime
\(long)
$$
$$
=-{K\over 2}
\int
\partial_{q_1}J(R) (z^\prime -z )
U (z^\prime) q_1 \text{d} q
\text {d} z^\prime
$$
$$
={K\over 2}
\int J(R) (z^\prime -z) U (z^\prime)
\text{d}q \text{d}z^\prime
$$
$$
={K\over 2}
\int (z^\prime -z)\tilde{J}
(z^\prime -z) U(z^\prime )\text{d}z^\prime
$$
in which the first equality follows because $J$ is
rotation invariant (isotropy), the second is integration
by parts with respect to $z^{\prime}$ and the fact that
$K=\sum_i k_i$, the third is the chain rule of differentiation, the fourth
is integration by parts with respect to $q_1$ and the last one follows from
the definition \(Jtilda).

This ends the verification of Statement 2.
\qed

\head
Appendix A: the surface tension.
\endhead
\resetall
\sectno=6

This Appendix generalizes [\rcite{Butta}], which
deals with the case of $f$  given in \(entropy). We are in
the same framework as Proposition 4.2.
The surface tension can be defined as the difference
 between the
free energy of an equilibrium state with
an interface and a homogeneous one
[\rcite{Butta},\rcite{Spohninterface}], i.e. from \(freeE)
$$
S=\lim_{L\rightarrow \infty}
{1\over (2L)^{d-1}} \lim_{M\rightarrow \infty}
\int_{-L}^{+L} \text{d} x_1 \ldots
\int_{-L}^{+L} \text{d} x_{d-1}
\int_{-M}^{+M} \text{d}x_d
\left[ g (\rho^*) (x)- f_{c}(\rho^+)\right]
\(Butta14)
$$
in which $\rho^*(x)=U (e_d \cdot x)$,
where $e_d=(0,\ldots ,0 , 1)\in \real ^d$, and
for any measurable function $\rho : \real^d \rightarrow [0,1]$
$$
g(\rho)(x)=f(\rho(x)) -{1\over 2} [\rho (x)-1/2]
(J*[\rho-1/2])(x)
\(appg)
$$
Observe that $f_{c}(\rho^+)=f_{c}(\rho^-)=g(\rho^+)=g(\rho^-)$.
\(Butta14) is well defined because of Proposition 4.2
and the properties of $f$.
Clearly \(Butta14) reduces to
$$
S=\int_{\real} \left[ g(\rho^*)(0,\ldots, 0,z)-f_{c}(\rho ^+)\right]
\text{d}z.
\(butta14r)
$$

We are going to  show that $S$, as defined in
\(Butta14), coincides with \(surftens).

Observe that for $\rho \in C^1$
$$
{\partial g \over \partial x_d}(\rho)=
f^\prime (\rho) {\partial\rho \over \partial x_d} -
{1\over 2} {\partial \rho \over \partial x_d} J*[\rho-1/2]
-{1\over 2}[\rho-1/2] J*{\partial \rho \over \partial x_d}
\(appint)
$$
but, by \(functional), $f^\prime (\rho^*)=
J* (\rho^*-1/2)$.
Hence
$$
\align
{\partial g \over \partial x_d}(\rho^*)&=
{1\over 2} {\partial \rho^* \over \partial x_d} J*[\rho^*-1/2]
-{1\over 2}[\rho^*-1/2] J*{\partial \rho^* \over \partial x_d}
\(appint2)
\\
&= {1\over 2}
\left(
U^\prime (x_d) (\tJ \star [U-1/2])(x_d) -
[U (x_d)-1/2] (\tJ \star U^\prime)(x_d) \right)
\endalign
$$
Integrating  \(butta14r) by parts and substituting
\(appint2) in the integrand
we finally obtain
$$
\align
S&=-{1\over 2} \int_\real x_d {\partial g \over \partial x_d}(\rho^*)
\text{d}x_d\\
&={1\over 2} \int_\real \int_\real
(z^\prime -z) U ^\prime (z) \tJ (z-z^\prime)
[U (z^\prime)-1/2]
\text{d}z \text{d} z^\prime\\
&= {1\over 2} \int_\real \int_\real
(z^\prime -z) U ^\prime (z) \tJ (z-z^\prime)
U (z^\prime)
\text{d}z \text{d} z^\prime
\(appfinal)
\endalign
$$
where we used
the fact that there is  an $m^\prime>0$ such that
$\lim_{z \rightarrow \pm \infty}
e^{m^\prime \vert z \vert}
U^\prime (z)=0$,
which follows from \(functional) and
Proposition 4.2, point (3). Since \(appfinal)
 is \(surftens), we are done.

\head
Ackowledgements
\endhead

It is a pleasure to acknoweledge several fruitful
discussions with P. Butt\`a, E. Orlandi, R.V. Kohn,
E. Presutti, L. Triolo and H. Spohn.
G.G. would like to thank B. Brighi and M.~Chipot
for several discussions and considerable help
on the material in Section 4.
This work was partially supported
by the Swiss National Science Foundation, Project  20--$41^\prime925.94$
 (G.G.) and by NSF--DMR
92--13424 4--20946 and AFOSR 0159 4--26435 (J.L.L.).

\enddocument